\documentclass[article,twocolumn,showpacs,preprintnumbers,amsmath,amssymb]{revtex4}
\usepackage[dvips]{color}
\usepackage{graphicx}
\usepackage{amsmath}
\usepackage{booktabs}
\begin{document}
\title{Lower bound of assortativity coefficient in scale-free networks}


\author{Dan Yang$^{1}$}
\author{Liming Pan$^{1}$}
\author{Tao Zhou$^{1,2}$}
\email{zhutou@ustc.edu}
\affiliation{
$^1$CompleX Lab, Web Sciences Center, University of Electronic Science and Technology of China,
Chengdu 611731, People's Republic of China\\
$^2$Big Data Research Centre, University of Electronic Science and Technology of China,
Chengdu 611731, People's Republic of China\\
}

\date{\today}

\begin{abstract}

The degree-degree correlation is important in understanding the structural organization of a network and the dynamics upon a network. Such correlation is usually measured by the assortativity coefficient $r$, with natural bounds $r \in [-1,1]$. For scale-free networks with power-law degree distribution $p(k) \sim k^{-\gamma}$, we analytically obtain the lower bound of assortativity coefficient in the limit of large network size, which is not -1 but dependent on the power-law exponent $\gamma$. This work challenges the validation of assortativity coefficient in heterogeneous networks, suggesting that one cannot judge whether a network is positively or negatively correlated just by looking at its assortativity coefficient.

\end{abstract}

\pacs{89.20.Hh, 89.20.Ff, 89.65.-s, 89.75.Fb}

\maketitle

\section{Introduction}

The last decade has witnessed a great change where the studies on networks, being of very limited interests mainly from mathematical society in the past, have received a huge amount of attention from many branches of sciences \cite{Newman2010,Chen2012}. The fundamental reason is that the network structure could well describe the interacting pattern of individual elements, which leads to many complex phenomena in biological, social, economic, communication, transportation and physical systems. Characterizing the structural features \cite{Costa2007,Barthelemy2011,Holme2012} is the foundation of the correct understanding about the dynamics of networks \cite{Albert2002,Dorogovtsev2002,Newman2003,Boccaletti2006}, the dynamics on networks (e.g., epidemic spreading \cite{Zhou2006,Pastor-Satorras2015}, transportation \cite{Tadic2007,Wang2007}, evolutionary game \cite{Szabo2007,Perc2010}, synchronization \cite{Arenas2008}, and other social and physical processes \cite{Dorogovtsev2008,Castellano2009}), as well as the network-related applications \cite{Fortunato2010,Costa2011,Lu2011,Lu2012}.

Thus far, the characterization of networks is still a challenging task, because some seemingly standalone structural properties are indeed statistically dependent to each other, resulting in many non-trivial structural constrains that are rarely understood \cite{Zhou2007,Lopez2008,Baek2012,Orsini2015}. Let's look at the five fundamental structural features of general networks: (i) the density $\rho$, quantified by the ratio of the number of edges $M$ to the possibly maximum value $N(N-1)/2$, where $N$ is the number of nodes; (ii) the degree $k_i$ of a node $i$ (defined as the number of $i$'s associated edges) as well as the degree distribution $p(k)$ \cite{Barabasi1999}; (iii) the average distance $\langle d \rangle$ over all pairs of nodes in a connected network \cite{Watts1998}; (iv) the assortativity coefficient $r$ that quantifies the degree-degree correlation (with mathematical definition shown later) \cite{Newman2002,Newman2003b}; (v) the clustering coefficient $c_i$ of a node (defined as the ratio of the number of edges between $i$'s neighbors to the possibly maximum value) and the average clustering coefficient $\langle c \rangle$ over all nodes with degree larger than 1 \cite{Watts1998}. Even though these measures are very simple compared with many other recently proposed network measures, there are complicated correlations between them. For example, scale-free networks are of very low density \cite{Genio2011} and usually small average distance \cite{Cohen2003}, a node of larger degree is often with smaller clustering coefficient \cite{Ravasz2003,Zhou2005}, a very high-density network must be with high clustering coefficient and small-world property \cite{Markov2013,Orsini2015}, networks with high clustering coefficients tend to have high assortativity coefficients \cite{Foster2011}, to name just a few.

In this paper, we focus on the mathematical relationship between assortativity coefficient and degree distribution. In particular, as the majority of real-world networks are found to be very heterogeneous in degree, we consider a typical class of networks \cite{Caldarelli2007,Barabasi2009}: the scale-free networks with power-law degree distribution $p(k)\sim k^{-\gamma}$, where $\gamma>0$ is called the power-law exponent. Both the degree distribution and assortativity coefficient are shown to be critical in understanding the structural organization of a network and the dynamics upon a network \cite{Dorogovtsev2002,Albert2002,Newman2003,Boccaletti2006,Caldarelli2007,Barabasi2009,Barrat2008,Noldus2015}, hence to uncover the correlation between these two fundamental measures could help us in clarifying whether a structural property is significant or just a statistical consequence of another property \cite{Orsini2015}.

Menche \emph{et al.} \cite{Menche2010} analyzed the maximally disassortative scale-free networks and found that the lower bound of assortativity coefficient, $r_{\min}$, will approach to zero when $2<\gamma<4$ as the increase of the network size $N$. Instead of an explicit value, they only provide the order of $r_{\min}$ in the large $N$ limit. Dorogovtsev \emph{et al.} \cite{Dorogovtsev2010} considered a specific class of recursive trees with power-law degree distribution and found that the assortativity coefficient is always zero. Raschke \emph{et al.} \cite{Raschke2010} showed both analytical and numerical results that the assortativity coefficient depends on the network size $N$. Litvak and Van Der Hofstad \cite{Litvak2013,Hofstad2014} highlighted the problem that the assortativity coefficient in disassortative networks systematically decreases with the network size, and they provide some mathematical explanation on this phenomenon. In particular, they showed that in the large $N$ limit, the assortativity coefficient is no less than zero \cite{Hofstad2014}. The above analyses, either on network configuration model with given degree distribution or on some specific scale-free network models, showed non-trivial dependencies between $r$ and $N$, as well as between $r$ and $\gamma$. However, to our best knowledge, an explicit lower bound of $r$ in scale-free networks has not been reported in the literatures. In this paper, we will derive the lower bound of $r$ in scale-free networks for $N\rightarrow \infty$, and then most of the above-mentioned conclusions can be considered as direct deductions from our results.

This paper is organized as follows. In the next section, we will analytically derive the lower bound of assortativity coefficient. In section III and section IV, we will validate the theoretic bound via simulation and empirical analysis, respectively. Lastly, we will draw the conclusion and discuss the theoretical significance and practical relevance of our contribution in section V.

\section{Theoretic Bound}

A network is called assortative if nodes tend to connect with other nodes with similar degrees, and disassortative if high-degree nodes tend to connect with low-degree nodes. In this paper, we focus on undirected simple networks, of which the degree correlation is usually measured by the assortativity coefficient, defined as~\cite{Newman2002,Newman2003b}:
\begin{equation}\label{eq:pearson}
  r=\frac{{{M}^{-1}}\sum\nolimits_{i}{{{j}_{i}}{{k}_{i}}}-{{[{{M}^{-1}}\sum\nolimits_{i}{\frac{1}{2}({{j}_{i}}+{{k}_{i}})}]}^{2}}}{{{M}^{-1}}\sum\nolimits_{i}{\frac{1}{2}({{j}_{i}}^{2}+{{k}_{i}}^{2})}-{{[{{M}^{-1}}\sum\nolimits_{i}{\frac{1}{2}({{j}_{i}}+{{k}_{i}})}]}^{2}}},
\end{equation}
where $M$ is the number of edges and ${{j}_{i}}$, ${{k}_{i}}$ are the degrees of the nodes at the ends of the $i$th edge, with $i=1,2,...,M$. The assortativity coefficient $r$ is actually the Pearson correlation coefficient between the degrees of neighboring nodes, which is supposed to have natural bounds $r \in [-1,1]$, where $r=-1$ indicates the completely negative correlation, while $r=+1$ suggests the perfectly positive correlation. It is straightforward to accept that a network is assortative when $r>0$ and disassortative when $r<0$ \cite{Newman2002,Newman2003b,Newman2003c}. However, such affirm could be wrong since the natural bounds do not imply a uniform distribution of $r$ in $[-1,1]$ and furthermore, the bounds of $r$ in a given network ensemble can be different from $[-1,1]$. For example, if for a given network ensemble, $r$ has non-trivial bounds $[0,1]$, claiming a network with a small positive $r$ to be assortative can be unreliable since it's quite close to the lower bound of all possible values. In such case, a network with zero or very small assortativity coefficient could be disassortative or assortative. In fact, Dorogovtsev \emph{et al.} \cite{Dorogovtsev2010} found a specific class of growing trees, which are strongly correlated but the assortativity coefficients are always zero.

\begin{figure}[!htb]
  \centering
  \includegraphics[width=0.48\textwidth]{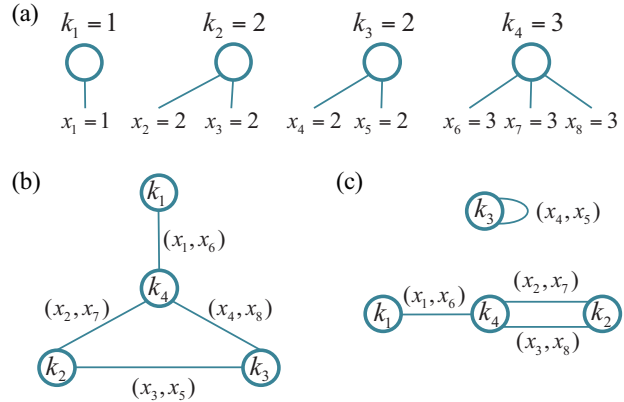}
  \caption{(Color online) Generating networks from a given degree sequence $\underline{k_{0}}=\{k_{1},k_{2},k_{3},k_{4}\}=\{1,2,2,3\}$. (a) Each stub is labeled by the degree of the node it is attached, and thus the stub sequence is obtained as $\underline{x_{0}}=\{x_{1},x_{2},\cdot\cdot\cdot,x_{7},x_{8}\}=\{1,2,2,2,2,3,3,3\}$.
  Realizations from the network ensemble defined by the degree sequence $\underline{k_{0}}$ can be represented by the corresponding $\underline{x_{0}}$, such as
   (b) $g_{\theta_{0}}=\{(x_{1},x_{6}),(x_{2},x_{7}),(x_{3},x_{5}),(x_{4},x_{8})\}$, one possible realization of simple networks without self-loops or multiple edges; and (c) $g_{\theta_{1}}=\{(x_{1},x_{6}),(x_{2},x_{7}),(x_{3},x_{8}),(x_{4},x_{5})\}$, one possible realization of networks with self-loops and multiple edges. }
  \label{illustrate}
\end{figure}

To uncover the non-trivial correlation between assortativity coefficient and degree distribution, we consider the most widely studied network ensemble, where the degree sequence is fixed and networks satisfying the degree constrain appear with equal probabilities~\cite{JPark2004,Bianconi2008}. Networks in such ensemble can be sampled by the uniform configuration model (UCM)~\cite{Newman2001}. Initially, each node $i$ is assigned a given degree $k_i$ according to the degree sequence and thus attached by $k_i$ half-edges (`stubs'). Then, random pairs of stubs are connected to form edges, without any self-loops and multiple edges. Considering a degree sequence $\underline{k}=\{k_{1},k_{2},k_{3},\cdot\cdot\cdot,k_{N}\}$ with $k_i$ being arranged in a non-decreasing order as $k_{1}\leq k_{2}\leq \cdot\cdot\cdot \leq k_{N}$. Obviously, there are $\sum^N_i k_i = 2M$ stubs. By labeling each stub with the degree of the attached node, one can obtain the stub sequence $\underline{x}=\{x_{1},x_{2},x_{3},\cdot\cdot\cdot,x_{2M}\}$, which is also arranged in a non-decreasing order as $x_{1}\leq x_{2}\leq \cdot\cdot\cdot \leq x_{2M}$. This procedure is illustrated in Fig.~\ref{illustrate}(a). The four circles denote four nodes with degrees $\underline{k_{0}}=\{k_{1},k_{2},k_{3},k_{4}\}=\{1,2,2,3\}$. The stubs are labeled with the degrees of nodes they are attached to, as $\underline{x_{0}}=\{x_{1},x_{2},\cdot\cdot\cdot,x_{7},x_{8}\}=\{1,2,2,2,2,3,3,3\}$.

It is clear that a possible realization of networks from the degree sequence $\underline{k}$ can also be represented by the corresponding stub sequence $\underline{x}$ as:
$g_{\theta}=\{(x_{\theta_{1}},x_{\theta_{2}}),(x_{\theta_{3}},x_{\theta_{4}}),(x_{\theta_{5}},x_{\theta_{6}}),\cdot\cdot\cdot,(x_{\theta_{2M-1}},x_{\theta_{2M}})\}$,
where $\theta$ is a rearrangement of the sequence $\{1,2,3,\cdot\cdot\cdot,2M\}$, and $\theta_{i}$ is the $i$-th element in the rearranged sequence. Consider then the degree sequence $\underline{k_{0}}$ in Fig.~\ref{illustrate}(a), by rearranging $\underline{x_{0}}$, we can get all possible networks generated from $\underline{k_{0}}$. As an illustration, Fig.~\ref{illustrate}(b) shows a possible realization of the network ensemble by $\underline{k_{0}}$: $g_{\theta_{0}}=\{(x_{1},x_{6}),(x_{2},x_{7}),(x_{3},x_{5}),(x_{4},x_{8})\}$.

For each $g_{\theta}$, the corresponding assortativity coefficient $r_{\theta}$ can be rewritten via the $\underline{x}$ sequence as:
\begin{eqnarray} \label{eq:rtheta}
  r_\theta =  \frac{{M^{-1}\sum_{i=1}^M x_{\theta_{2i-1}}x_{\theta_{2i}} - {\langle x \rangle}^{2}}}{\langle x^{2}\rangle-\langle x\rangle^{2}},
\end{eqnarray}
where
\begin{equation}
\langle x \rangle =\frac{1}{M}\sum_{i=1}^M(x_{\theta_{2i-1}}\!\!\!+\!\!x_{\theta_{2i}})
\end{equation}
and
\begin{equation}
\langle x^{2} \rangle = \frac{1}{M}\sum_{i=1}^M(x^{2}_{\theta_{2i-1}}\!\!\!+\!\!x^{2}_{\theta_{2i}}).
\end{equation}
From Eq.~(\ref{eq:rtheta}), it is observed that the first term in the dominator
\begin{equation}
S_{\theta}=\frac{1}{M}\sum_{i=1}^M x_{\theta_{2i-1}}x_{\theta_{2i}}
\end{equation}
is the only term that the rearrangement $g_{\theta}$ affects, while other terms are fixed given $\underline{x}$. That is to say, $g_{\theta_{\mathrm{min}}}$ which minimize $S_{\theta}$ will also minimize the assortativity coefficient $r$. Actually $g_{\theta_{\mathrm{min}}}$ can be determined based on the branch-and-bound idea~\cite{Hallin1992,Guo2015}, which is $g_{\theta_{\mathrm{min}}}=\{(x_{1},x_{2M}),(x_{2},x_{2M-1}),(x_{3},x_{2M-2}),\cdot\cdot\cdot,(x_{M},x_{M+1})\}$ (see \emph{Appendix A} for the proof of the above proposition).

Notice that, in general $g_{\theta}$ can not guarantee the absence of self-loops and multiple edges. For example, as showed in Fig.~\ref{illustrate}(c), $g_{\theta_{1}}=\{(x_{1},x_{6}),(x_{2},x_{7}),(x_{3},x_{8}),(x_{4},x_{5})\}$ is one possible realization in the network ensemble $g_\theta$, which has both one self-loop and one multiple edge. However we can prove that if $p(k)\sim k^{-\gamma}$ with $\gamma >2$, the self-loops and multiple edges will vanish in the thermodynamical limit (i.e., $N\rightarrow \infty$) for the specific network generated by $g_{\theta_{\mathrm{min}}}$ (see \emph{Appendix B} for the proof).

Taking $g_{\theta_{\mathrm{min}}}$ into Eq.~(\ref{eq:rtheta}), the lower bound of the assortativity coefficient $r$ is:
\begin{equation}\label{eq:rmin}
  r_{{\mathrm{min}}}=\frac{{M^{-1}\sum_{i=1}^M x_{i}x_{2M+1-i}  - {\langle x\rangle}^{2}}}{\langle x^{2}\rangle-\langle x\rangle^{2}},
\end{equation}
which holds for any fixed degree sequence that allows $g_{\theta_{\mathrm{min}}}$ to be a simple network without any self-loops or multiple edges. Now we consider the expected value $\mathrm{E}[r_{\mathrm{min}}]$ when the degree sequence is drawn from a power-law distribution $p(k)\sim k^{-\gamma}$. Del Genio \emph{et al.}~\cite{Genio2011} showed a fundamental mathematical constraint that when $0 \leq \gamma \leq 2$, the graphical fraction (i.e., the ratio of the number of graphical sequences that can be realized as simple networks to the total number of degree sequences with an even degree sum in a given ensemble) approaches zero, and in fact the majority of real scale-free networks are of power-law exponents $\gamma>2$~\cite{Caldarelli2007,Clauset2009}. Therefore, in this work, we focus on scale-free networks with $\gamma>2$. Obviously, if the degrees $\underline{k}$ are sampled from a power-law distribution $p(k)\sim k^{-\gamma}$, the corresponding stubs $\underline{x}$ also follow a power-law distribution $f(x)\sim x^{-\beta}$ with $\beta=\gamma-1$.

\begin{figure}[!htb]
  \centering
  \includegraphics[width=0.48\textwidth]{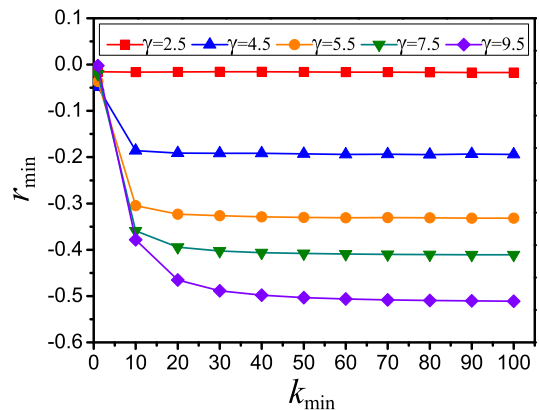}
  \caption{(Color online) Simulated results of $r_{\mathrm{min}}$ versus $k_{\mathrm{min}}$ for different degree exponents $\gamma$. The five curve from up to bottom denote the sample average with $\gamma =2.5$, $\gamma =4.5$, $\gamma =5.5$, $\gamma =7.5$ and $\gamma =9.5$, respectively. The network size is fixed as $N=10^6$. Each data point is obtained by averaging over 2000 independent runs. It is observed that $r_{\mathrm{min}}$ converges as the increase of $k_{\mathrm{min}}$. }
  \label{RminKm}
\end{figure}

As $\underline{x}$ is arranged in non-decreasing order, we have $x_i\leq x_M, \forall i \leq M$. Thus the following inequality always holds:
\begin{equation}\label{eq:ineq1}
\begin{split}
{{S}_{\theta}}=&{{M}^{-1}}\sum\nolimits_{i=1}^{M}{{{x}_{i}}{{x}_{2M-1+i}}} \\
\le & {{x}_{M}}{{M}^{-1}}\sum\nolimits_{i=1}^{M}{{{x}_{2M-1+i}}}.
\end{split}
\end{equation}
When $N\rightarrow\infty$, the right hand side of Eq.~(\ref{eq:ineq1}) can be approximated by continuous variables as:
\begin{eqnarray}\label{eq:eqnar1}
  {{x}_{M}}{{M}^{-1}}\sum\nolimits_{i=1}^{M}{{{x}_{2M+1-i}}}\approx 2{{x}_{M}}\int_{{{x}_{M}}}^{N-1}{xf(x)dx}.
\end{eqnarray}
Doing the integral in Eq.~(\ref{eq:eqnar1}), and taking into account the fact that $x_{M}\sim M^{0}$~\cite{Baek2012}, we have:
\begin{eqnarray}\label{eq:eqnar2}
&S_{\theta}=&\left\{ \begin{matrix}
   O\left( {{N}^{3-\gamma }} \right)~~~~2<\gamma <3 \\
   O\left( \ln N \right)~~~~~~~~~~~~\gamma=3 \\
   O\left( 1 \right)~~~~~~~~~~3<\gamma <4.  \\
\end{matrix} \right
.
\end{eqnarray}
Similarly, we also have:
\begin{equation}\label{eq:x}
  \langle x\rangle=\left\{ \begin{matrix}
   O\left( {{N}^{3-\gamma }} \right)~~~~2<\gamma <3 \\
   O\left( \ln N \right)~~~~~~~~~~~~\gamma=3 \\
   O\left( 1 \right)~~~~~~~~~~3<\gamma <4  \\
\end{matrix} \right
.
\end{equation}
and
\begin{equation}\label{eq:x}
  \langle x^2 \rangle=\left\{ \begin{matrix}
   O\left( {{N}^{4-\gamma }} \right)~~~~2<\gamma <4 \\
   O\left( \ln N \right)~~~~~~~~~~~~\gamma=4. \\
\end{matrix} \right
.
\end{equation}
Evidently, the denominator has the strict larger order when $2<\gamma \leq 4$, hence
\begin{equation}\label{eq:Rmin0}
  r_{\mathrm{min}}=0
\end{equation}
in the thermodynamic limit $N\rightarrow \infty$.

In the case $\gamma>4$, we show numerically that $r_{\mathrm{min}}$ decreases with $k_{\mathrm{min}}$ and eventually approaches to a steady value as $k_{\mathrm{min}}$ becomes large, as shown in Fig.~\ref{RminKm}. Therefore we set a relatively large $k_{\mathrm{min}}$, which also allows us to take variables in $\underline{x}$ as continuous. Below, we carry out our derivation based on continuous variables, thus the distribution function of $\underline{x}$ can be normalized as
\begin{equation}
  \tilde{f}(x)=\frac{\beta -1}{{{k}_{\mathrm{min}}}}{{(\frac{x}{{{k}_{\mathrm{min}}}})}^{-\beta }}
\end{equation}

For $\gamma>4$, both of the first or second moment of $\underline{x}$ exsit, and they can be calculated as follows,
\begin{equation}
  \langle x\rangle =\int_{{{k}_{\mathrm{min}}}}^{N-1}{x{\tilde{f}}\,}(x)dx=\frac{\gamma -2}{\gamma -3}{{k}_{\mathrm{min}}}
\end{equation}
and
\begin{equation}
  \langle x^2\rangle =\int_{{{k}_{\mathrm{min}}}}^{N-1}{x^{2}{\tilde{f}}\,}(x)dx=\frac{\gamma -2}{\gamma -4}{k}_{\mathrm{min}}^{2}
\end{equation}
when $N\rightarrow\infty$.

\begin{figure}[!h]
  \centering
  \includegraphics[width=0.48\textwidth]{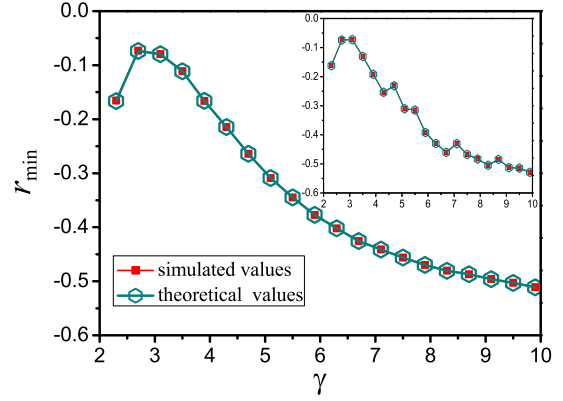}
  \caption{(Color online) Simulated results of $r_{\mathrm{min}}$ versus the theoretical prediction by Eq. (6). The blue hexagons represent the theoretical values of the lower bound of assortativity coefficients of a series of randomly generated scale-free networks with different degree exponents. The red squares denote the simulated values, obtained by the degree-preserving edge-rewiring procedure. The network size is set to be $N=50000$ and the minimum degree is set as $k_{\min}=50$. Each data point is averaged over 500 independent runs. The inset shows the results of a single run, where the simulation result is still perfectly in accordance with the theoretical prediction.} \label{fig2}
\end{figure}

We divide each element in $\underline{x}$ by ${k}_{\mathrm{min}}$, then we get a new sequence $\underline{\hat{x}}$, which certainly follows the power-law distribution with the same exponent $\beta$, but the minimum value is rescaled to be $1$, i.e., $\hat{f}(x)=(\beta-1)x^{-\beta}$. Obviously
\begin{equation}\label{SS}
  {S}_{\theta_{\mathrm{min}}}={k}_{\mathrm{min}}^{2}\hat{S}_{\theta_{\mathrm{min}}}.
\end{equation}
To get the expected value of $\hat{S}_{\theta_{\mathrm{min}}}$, we introduce the joint distribution of order statistics $p\left( {{\hat{x}}_{m}}={{t}_{1}},{{\hat{x}}_{n}}={{t}_{2}} \right)$,
which is the probability that the $m$th element in the $\underline{\hat{x}}$ takes value $t_1$, while the $n$th takes value $t_2$. It reads~\cite{David2003}
\begin{equation}
\begin{split}
 p&\left( {{\hat{x}}_{m}}={{t}_{1}},{{\hat{x}}_{n}}={{t}_{2}} \right)=(2M)!\frac{{{\left[ F\left( {{t}_{1}} \right) \right]}^{m-1}}}{\left( m-1 \right)!}\\
 \times& \frac{{{\left[ F\left( {{t}_{2}} \right)-F\left( {{t}_{1}} \right) \right]}^{n-m-1}}}{\left( n-m-1 \right)!}\\
 \times& \frac{{{\left[ 1-F\left( {{t}_{2}} \right) \right]}^{2M-n}}}{\left( 2M-n \right)!}\hat{f}\left( {{t}_{1}} \right)\hat{f}\left( {{t}_{2}} \right),
\end{split}
\end{equation}
where $F(t)$ is the cumulative distribution function
\begin{equation}
  F(t)=\int_{1}^{t}{\hat{f}(x)dx}=1-t^{1-\beta}.
\end{equation}

Using a shorthand $c=\frac{1}{\beta-1}$, we express $\mathrm{E}\left[ {{\hat{x}}_{i}}{{\hat{x}}_{2M+1-i}} \right]$ in terms of $p\left( {{\hat{x}}_{m}}={{t}_{1}},{{\hat{x}}_{n}}={{t}_{2}} \right)$ as:
\begin{equation}
\begin{split}
  \mathrm{E}&\left[ {{\hat{x}}_{i}}{{\hat{x}}_{2M+1-i}} \right]\!\!\\
  =&\!\!\!\int\!\!\!\!\int_{1\leq {{t}_{1}}\leq {{t}_{2}}}^{{\infty }}\!\!\!\!\,{{t}_{1}}{{t}_{2}}p\left( {{\hat{x}}_{i}}={{t}_{1}},{{\hat{x}}_{2M+1-i}}={{t}_{2}} \right)d{{t}_{1}}d{{t}_{2}} \\
  =&\frac{\Gamma\left( 2M+1 \right)}{\Gamma\left( 2M+1-2c \right)}\frac{\Gamma\left( i+2-c \right)}{\Gamma\left( i+2 \right)}\frac{\Gamma\left( 2M-i+1-2c \right)}{\Gamma\left( 2M-i+1-c \right)} \\
  =&{{\left( 2M+1 \right)}^{2c}}{{\left( i+2-c \right)}^{-c}}{{\left( 2M-i+1-2c \right)}^{-c}},
\end{split}
\end{equation}
where $\Gamma(x)=\int_0^{\infty} t^{x-1}e^{-t}dt$ is the well-known Gamma function and $\Gamma(n)=(n-1)!$ for any positive integer $n$.

Obviously, $M\rightarrow \infty$ in the thermodynamic limit $N \rightarrow \infty$, and then
\begin{equation}
\begin{split}
  \hat{S}_{\theta_{\mathrm{min}}}=&\underset{M\to \infty }{\mathop{\lim }}\,{{M}^{-1}}\sum\nolimits_{i=1}^{M}\mathrm{E}\left[ {{\hat{x}}_{i}}{{\hat{x}}_{2M+1-i}} \right] \\
  =&2\underset{M\to \infty }{\mathop{\text{lim}}}\,\frac{\sum\nolimits_{i=1}^{M}\,{{\left( i+2-c \right)}^{-c}}{{\left( 2M-i+1-2c \right)}^{-c}}}{{{\left( 2M+1 \right)}^{1-2c}}{}} \\
  =&2\underset{M\to \infty }{\mathop{\text{lim}}}\,\frac{\sum\nolimits_{i=1}^{M}\,{{\left( \frac{i}{2M+1} \right)}^{-c}}{{\left( 1-\frac{i}{2M+1} \right)}^{-c}}}{{{\left( 2M+1 \right)}}{}} \\
  =&2\int_{1}^{\frac{1}{2}}{{u}^{-c}}{{\left( 1-u \right)}^{-c}}du \\
  =&2B\left( \frac{1}{2};\frac{\beta -2}{\beta -1},\frac{\beta -2}{\beta -1} \right),
\end{split}
\end{equation}
where $u=\frac{i}{2M+1}$ and $B(\cdot)$ is the \emph{incomplete beta function}
\begin{equation}
	B(x; a,b) = \int_{0}^{x} t^{a-1} (1-t)^{b-1} dt.
\end{equation}
According to Eq.~(\ref{SS}), we thus get
\begin{equation}\label{eq:S}
  {S}_{\theta_{\mathrm{min}}}=2{k}_{\mathrm{min}}^{2}B\left( \frac{1}{2};\frac{\beta -2}{\beta -1},\frac{\beta -2}{\beta -1} \right).
\end{equation}
When $\gamma>4$, substituting Eq.~(\ref{eq:S}) into Eq.~(\ref{eq:rmin}), we have
\begin{equation} \label{eq:gamma4}
\begin{split}
  {{r}_{\mathrm{min}}}=&\frac{{{M}^{-1}}\mathop{\sum }_{i=1}^{M}{{x}_{i}}{{x}_{2M+1-i}}-{{\left\langle x \right\rangle }^{2}}}{\left\langle {{x}^{2}} \right\rangle -{{\left\langle x \right\rangle }^{2}}} \\
  =&\frac{2B\left( \frac{1}{2};\frac{\beta -2}{\beta -1},\frac{\beta -2}{\beta -1} \right)-{{\left( \frac{\beta -1}{\beta -2} \right)}^{2}}}{\frac{\beta -1}{\beta -3}-{{\left( \frac{\beta -1}{\beta -2} \right)}^{2}}}\\
  =&\frac{2B\left( \frac{1}{2};\frac{\text{ }\!\!\gamma\!\!\text{ }-3}{\text{ }\!\!\gamma\!\!\text{ }-2},\frac{\text{ }\!\!\gamma\!\!\text{ }-3}{\text{ }\!\!\gamma\!\!\text{ }-2} \right)-{{\left( \frac{\text{ }\!\!\gamma\!\!\text{ }-2}{\text{ }\!\!\gamma\!\!\text{ }-3} \right)}^{2}}}{\frac{\text{ }\!\!\gamma\!\!\text{ }-2}{\text{ }\!\!\gamma\!\!\text{ }-4}-{{\left( \frac{\text{ }\!\!\gamma\!\!\text{ }-2}{\text{ }\!\!\gamma\!\!\text{ }-3} \right)}^{2}}}.
\end{split}
\end{equation}
Combining Eq.~(\ref{eq:gamma4}) and Eq.~(\ref{eq:Rmin0}), the lower bound of scale-free networks is thus obtained as:
\begin{equation}
  {{r}_{\mathrm{min} }}=\left\{ \begin{matrix}
   0~~~~~~~~~~~~~~~~~~~~~~~~~~~~~~2<\gamma \le 4,~  \\
   \frac{2B\left( \frac{1}{2};\frac{\text{ }\!\!\gamma\!\!\text{ }-3}{\text{ }\!\!\gamma\!\!\text{ }-2},\frac{\text{ }\!\!\gamma\!\!\text{ }-3}{\text{ }\!\!\gamma\!\!\text{ }-2} \right)-{{\left( \frac{\text{ }\!\!\gamma\!\!\text{ }-2}{\text{ }\!\!\gamma\!\!\text{ }-3} \right)}^{2}}}{\frac{\text{ }\!\!\gamma\!\!\text{ }-2}{\text{ }\!\!\gamma\!\!\text{ }-4}-{{\left( \frac{\text{ }\!\!\gamma\!\!\text{ }-2}{\text{ }\!\!\gamma\!\!\text{ }-3} \right)}^{2}}}~~~~~~~~~~\gamma >4.  \\
\end{matrix}\text{ }\!\!~\!\!\text{  }\!\!~\!\!\text{  }\!\!~\!\!\text{  }\!\!~\!\!\text{ } \right.
\end{equation}

\begin{figure}[!ht]
  \centering
  \includegraphics[width=0.48\textwidth]{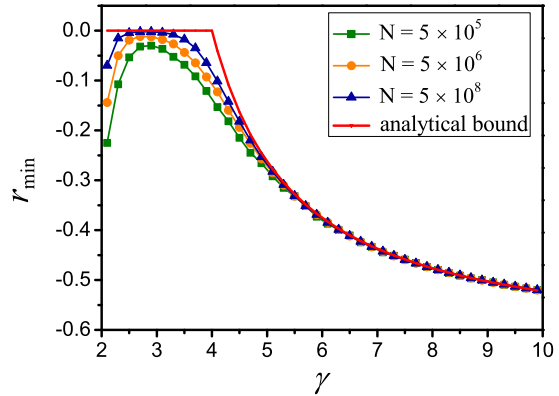}
  \caption{(Color online) The simulation result of the expected value of $r_{\mathrm{min}}$ for different $\gamma$, compared with the analytical result in Eq. (24). The red line represents the analytical result Eq. (24), and the squares, origin circle and blue up-triangles denote the simulation results for $N=10^{5}$, $N=10^{6}$, $N=10^{8}$, respectively. The minimum degree is fixed as $k_{\min}=150$. Each data point is averaged over 500 independent runs.} \label{fig3}
\end{figure}

\section{Simulations}

We verify the analytical results by extensive numerical simulations. Given the degree sequence, we search for the network with minimum assortativity (i.e., the most disassortative network) via a degree-preserving edge-rewiring procedure~\cite{Maslov2002,Kim2004,Zhao2006}. Specifically, to find the most disassortative network~\cite{Xulvi2004}, at each step two edges with four nodes at their ends are chosen at random. Now, we label these four nodes with $a$, $b$, $c$, and $d$ such that their degrees $k_a$, $k_b$, $k_c$ and $k_d$ are ordered as
\begin{equation}
k_a\geq k_b \geq k_c \geq k_d.
\end{equation}
Considering the operation to break the two edges and connect $a$ with $d$ and $b$ with $c$, if such operation will not generate multiple edges, we implement such operation, otherwise we do nothing. Since this edge-rewiring procedure has been proven to be ergodic~\cite{Maslov2004}, and $r$ is convex in the configuration space,
the procedure will eventually converge to the network with $r_{\mathrm{min}}$ (see more details in~\cite{Xulvi2004}).

In the simulation, we first generate the degree sequence by the distribution $p(k)\sim k^{-\gamma}$. Given $N$ and $k_{\min}$, we will sample $N$ integers between $k_{\min}$ and $N-1$ and check if these $N$ degrees obey the graphical condition according to the Erd\"os-Gallai theorem~\cite{Erdos1960}. If the degree sequence is graphical, it will be accepted and a scale-free network will be generated by the uniform configuration model~\cite{Newman2001}, otherwise it will be rejected. Notice that, when $\gamma>2$, almost every degree sequence is graphical~\cite{Genio2010}, and in fact we have not found any non-graphical degree sequence when $N\geq 5\times 10^4$ and $\gamma>2$ in the simualtion.

First of all, we compare the simulation results with Eq.~(\ref{eq:rmin}), which is the starting point of our analysis. As shown in figure 3, for both averaged results and single-run result, the theoretical prediction perfectly agrees with the simulation, demonstrating the validation of Eq.~(\ref{eq:rmin}). Figure 4 shows the expected value $r_{\min}$ for the scale-free networks with degree distribution $p(k)\sim k^{-\gamma}$, with $\gamma$ varying from 2 to 10. Overall speaking, the simulation results agree well with the analytical results, and with the increasing of the network size $N$, the simulation results are getting close to the theoretical bound. Notice that, nearing $\gamma=2$ and $\gamma=4$, the deviation between analytical and simulation results is larger, which is resulted from the fact that the order of divergence in the numerator and dominator of Eq.~(\ref{eq:rmin}) becomes close (see also Eqs. (9)-(11)).

\section{Empirical Evidence}
In this section, we test our theory on three large-scale real networks. (\romannumeral1) AS-Skitter~\cite{Leskovec2005}: Autonomous systems topology graph of the Internet. (\romannumeral2) YouTube~\cite{Leskovec2012}: The YouTube social network, in which users are connected with their friends in YouTube. (\romannumeral3) Web-Google~\cite{Leskovec2009}: Nodes represent web pages and edges represent hyperlinks between them, disgarding the direction of links. All the three networks are undirected simple networks without any self-loops or multiple edges. As shown in figure 5, these networks are all scale-free networks with $\gamma>2$.

\begin{figure}[!h]
  \centering
  \includegraphics[width=0.48\textwidth]{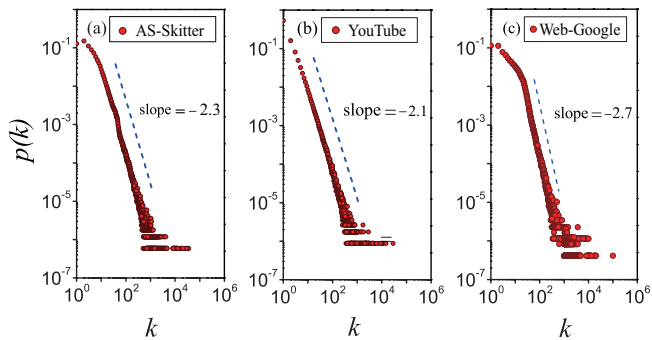}
  \caption{(Color online) Degree distribution of the three real networks: (a) AS-Skitter, (b) YouTube, and (c) Web-Google. All the three networks are scale-free networks with power-law exponents being 2.3, 2.1 and 2.7, respectively. The power-law exponents are estimated by using the maximum likelihood method~\cite{Clauset2009}. }
  \label{fig1}
\end{figure}

As shown in table 1, the assortativity coefficients of the three real networks are all very close to zero, in accordance with the theoretical prediction in Eq. (24). Considering the finite-size effects, as observed in figure 4, our theory agrees well with the empirical results. Furthermore, by using the degree-preserving edge-rewiring method, we can obtain the minimum and maximum assortativity coefficients. One can see from the last two columns in table 1 that the interval between $r_{\mathrm{min}}$ and $r_{\mathrm{max}}$ is quite narrow, similar to the phenomenon reported in Ref.~\cite{Zhou2007}. This phenomenon strongly suggests that the widely applied assortativity coefficient is not a suitable measure for the degree-degree correlation at least for scale-free networks, since to know the coefficient, we still cannot make sure whether this network is positively correlated or not.

\begin{table}
  \centering
  \begin{tabular}{ccccccc}
  \hline
  Network &N  &M  &$\gamma$ &$r$ & $r_{\mathrm{min}}$&$r_{\mathrm{max}}$ \\  \hline
  As-Skitter &1965206 &11095298 &2.3 &$-0.081$ & $-0.086$ &$-0.045$ \\
  YouTube &1134890 &2987624 &2.1 &$-0.037$ & $-0.044$ &$-0.004$ \\
  Web-Google &875731 &4322051 &2.7 &$-0.055$ & $-0.065$ &$0.108$ \\
  \hline
  \end{tabular}
  \caption{The basic statistics and assortativity coefficients of the three real networks. $N$ and $M$ are the number of nodes and the number of edges, and $\gamma$ is the estimated power-law exponents. $r$ is the assortativity coefficient, while $r_{\mathrm{min}}$ and $r_{\mathrm{max}}$ denote the minimum and maximum assortativity coefficients obtained by the degree-preserving edge-rewiring procedure. The method to obtain $r_{\mathrm{max}}$ is similar to that for $r_{\mathrm{min}}$. }
  \label{table1}
\end{table}

\section{Conclusion and Discussion}

This paper argued the invalidation of the well-known assortativity coefficient $r$ in highly heterogeneous networks, and in particular analytically obtained the lower bound of $r$ for scale-free networks with power-law exponent $\gamma >2$. According to the main result Eq. (24), when $2<\gamma \leq 4$, $r_{\min}$ will approach zero in the large $N$ limit and when $\gamma>4$, $r_{\min}$ will monotonously decrease as the increase of $\gamma$. In addition, as indicated by the simulation result in Figure 4, as the increase of network size, the lower bound $r_{\min}$ will also increase. The above results are in accordance with previous studies \cite{Menche2010,Dorogovtsev2010,Raschke2010,Litvak2013,Hofstad2014}, meanwhile the advantage of the present study is that it considered a more general case instead of specific network models and derived the explicit lower bound of $r$. At the same time, there are still some disadvantages in the present work. For example, we have not obtained the upper bound of $r$ or the analytical relation between $r_{\min}$ and $\gamma$ for finite-size networks. These problems may be solved in the future studies, but we do not know whether they can be solved under the present framework. In addition, the in-depth analyses on the degree-degree correlation in directed networks are very interesting and challenging, as such kind of correlation in directed networks is much more complicated~\cite{Williams2014,Hoom2015}.

In addition to the technical skills, this paper has demonstrated an important point of view that the assortativity coefficient is not a suitable measure for degree-degree correlation in heterogenous networks, since the possible range of $r$ in heterogeneous networks is very narrow~\cite{Zhou2007}, and to know the value of $r$ is usually not enough to draw a conclusion whether the target network is assortative or disassortative. Some scientists have suggested other measures for heterogeneous networks, most of which are rank-based coefficients, such as the Kendall-Gibbons' Tau~\cite{Kendall1990} suggested by Raschke \emph{et al.}~\cite{Raschke2010} and the Spearman's Rho~\cite{Spearman1904} suggested by Litvak and Van Der Hofstad \cite{Litvak2013,Hofstad2014}. However, compared with the extensive studies on assortativity coefficient~\cite{Noldus2015}, the studies on rank-based coefficients in complex networks are very limited. Although we still do not know whether a rank-based coefficient is the best candidate in properly measuring degree-degree correlation in heterogeneous networks, to uncover the statistical properties of rank-based coefficients and to explore other possible candidates based on the more in-depth understanding of the network ensemble are significant in the current stage.

\begin{acknowledgments}
    The authors acknowledge Ya-Jun Mao, Zhi-Hai Rong, Wei Wang, Yifan Wu and Zhi-Dan Zhao for valuable discussion and research assistance, and the Stanford Large Network Dataset Collection (SNAP Datasets) for the real data. This work is partially supported by National Natural Science Foundation of China under Grants Nos. 61433014 and 11222543.
\end{acknowledgments}

\begin{appendix}
\section{Determining $g_{\theta_{\mathrm{min}}}$ that minimizes $r$}

Inspired by the branch-and-bound idea~\cite{Hallin1992,Guo2015}, in this section we will prove that $g_{\theta_{\mathrm{min}}}$ minimizes $r$, where
$g_{\theta_{\mathrm{min}}}=\{(x_{1},x_{2M}),(x_{2},x_{2M-1}),(x_{3},x_{2M-2}),\cdot\cdot\cdot,(x_{M},x_{M+1})\}$.

Given the degree sequence and the corresponding ensemble, we define ${{\xi }^{\left( 0 \right)}}$ as the set of all possible realizations $g_{\theta}$, and at each step $\epsilon \geq 1$ we subdivide ${{\xi }^{\left( \epsilon-1 \right)}}$ into two nonempty subsets ${{\xi }^{\left( \epsilon \right)}}$ and ${{\xi }^{\left( \epsilon-1 \right)}}\backslash {{\xi }^{\left( \epsilon \right)}}$ according to the following rule, which guarantees that ${{\xi }^{\left( \epsilon \right)}}$ contains at least one realization $g_{\theta}$ that minimizes $r$.

In step one, we consider the stub of largest value and the stub of smallest value, which are $x_{2M}$ and $x_{1}$ respectively. If they are not connected, there must be two links $\left( {{x}_{1}},{{x}_{i}} \right)$ and $\left( {{x}_{j}},{{x}_{2M}} \right)$, with $1<i<2M$, $1<j<2M$ and $i\ne j$. Since $\underline{x}$ is arranged in a non-decreasing order, it is obvious that ${{x}_{1}}{{x}_{2M}}+{{x}_{i}}{{x}_{j}}\le {{x}_{1}}{{x}_{i}}+{{x}_{j}}{{x}_{2M}}$. Therefore, keeping other links the same, according to Eq. (2), the network with edges $\left( {{x}_{i}},{{x}_{j}} \right)$ and $\left( {{x}_{1}},{{x}_{2M}} \right)$ has no larger assortativity coefficient than the one with edges $\left( {{x}_{1}},{{x}_{i}} \right)$ and $\left( {{x}_{j}},{{x}_{2M}} \right)$. Then we define ${{\xi }^{\left( 1 \right)}}$ as the set of all realizations containing the edge $\left( {{x}_{1}},{{x}_{2M}} \right)$. Analogously, we further consider whether $x_{2M-1}$ and $x_{2}$ are connected in all realizations $g_{\theta} \in {{\xi }^{\left( 1 \right)}}$. Under the same arguments, ${{\xi }^{\left( 2\right)}}$ can be defined as the set of all realizations containing both edges $\left( {{x}_{1}},{{x}_{2M}} \right)$ and $\left( {{x}_{2}},{{x}_{2M-1}} \right)$.

After $M$ steps, the above process ends up with ${{\xi }^{\left( M \right)}}=\{{g_{\theta_{\mathrm{min}}}}\}$, that is
\begin{equation}\label{}
\begin{split}
  &{{\xi }^{\left( M\right)}}=\{{g_{\theta_{\mathrm{min}}}}\} \\
  &=\{(x_{1},x_{2M}),(x_{2},x_{2M-1}),\cdot\cdot\cdot,(x_{M},x_{M+1})\}.
\end{split}
\end{equation}

\section{Self-loops and multiple edges in $g_{\theta_{\mathrm{min}}}$ vanish in the thermodynamical limit}
The $g_{\theta_{\mathrm{min}}}$ derivated in \emph{Appendix A} can not guarantee the absence of self-loops or multiple edges. Here we show that actually self-loops and multiple edges
vanish in the thermodynamical limit for the specific structure of $g_{\theta_{\mathrm{min}}}$, given $p(k)\sim k^{-\gamma}$ with $\gamma >2$.

First let's consider self-loops. Clearly self-loops may appear only when two stubs of the same value are connected, since they might come from the same node. Under the specific arrangement of $g_{\theta_{\mathrm{min}}}$, stubs of values $\geq x_M$ always connect with stubs of values $\leq x_M$, thus connections among stubs with the same value only occur when their values are equal to $x_M$. Denote the number of stubs of value $x_M$ by $N(x_M)$ and the number of nodes with degree $x_M$ by $\widehat{N}(x_M)$, then it is obvious that
\begin{equation}
N(x_M)=x_M\times \widehat{N}(x_M).
\end{equation}
There are in total $\frac{1}{2} \widehat{N}(x_M) (\widehat{N}(x_M)-1)$ potential node pairs and in the worst case, $N(x_M)/2$ connections among these stubs will be generated. To avoid self-loops, there must be no less potential node pairs than the demanded connections, namely
\begin{equation}\label{eq:apB}
\frac{1}{2} \widehat{N}(x_M) (\widehat{N}(x_M)-1) \geq \frac{N(x_M)}{2}.
\end{equation}
Combining Eq. (B1) and Eq. (B2), the condition is reduced to $\widehat{N}(x_M) \geq x_M +1$. Actually for $x_M$ we have $x_M \sim {M^0}$, meanwhile $\widehat{N}(x_M)\sim N$~\cite{Baek2012,Genio2011}. Thus in the thermodynamical limit (i.e., $N\rightarrow \infty$), $\widehat{N}(x_M) \geq x_M +1$ holds. That is to say, the self-loops vanish in the thermodynamical limit.

Secondly, we consider the multiple edges. For an arbitrary node with degree $k$, which will draw $n (n \leq k)$ edges to the nodes with degree $k'$ according to the realization $g_{\theta_{\mathrm{min}}}$. Obviously, multiple edges can be avoided if $n \leq \widehat{N}(k')$, where $\widehat{N}(k')$ is the number of nodes with degree $k'$ in the given degree sequence. Thus if $n\leq \widehat{N}(k')$ holds for all possible values of $n$ and $k'$ in $g_{\theta_{\mathrm{min}}}$, we can conclude that multiple edges can be excluded for the certain realization $g_{\theta_{\mathrm{min}}}$. Notice that, in $g_{\theta_{\mathrm{min}}}$, two nodes are connected only if one is of degree $\geq x_M$ and the other is of degree $\leq x_M$. Therefore, we can only consider the case $k \geq x_M$ and $k' \leq x_M$, since if for all nodes with degree $k \geq x_M$, the multiple edges can be avoided, then for nodes with degree $k<x_M$, the multiple edges can also be avoided as these small-degree nodes cannot connect to each other in $g_{\theta_{\mathrm{min}}}$.

In the thermodynamical limit of scale-free networks with degree distribution $p(k)\sim k^{-\gamma}$, the maximum degree scales in the order $k_{\mathrm{max}} \sim N^{\frac{1}{\gamma-1}}$~\cite{Bogun¨¢2004}, hence when $\gamma >2$, we have
\begin{equation}
n\leq k \leq k_{\mathrm{max}} \sim N^{\frac{1}{\gamma-1}} < N
\end{equation}
for $N\rightarrow \infty$. At the same time, since $k' \leq x_M$, we have $\widehat{N}(k') \geq \widehat{N}(x_M)$, meanwhile $\widehat{N}(x_M)\sim N$ in the thermodynamical limit, that is
\begin{equation}
\widehat{N}(k') \geq \widehat{N}(x_M) \sim N.
\end{equation}
Combining Eq. (B3) and Eq. (B4), for all possible values of $n$ and $k'$, $n\leq \widehat{N}(k')$ in the limit $N\rightarrow \infty$. That is to say, the multiple edges vanish in the thermodynamical limit.

\end{appendix}

\end{document}